\documentstyle[NATO,epsf,namedreferences]{CRCKAPB}

\newcommand{\kms}{km\,s$^{-1}$}
\newcommand{\prcite}[2]{\protect\citeauthoryear{#1}{#2}}
\newcommand{\scite}[1]{\shortcite{#1}}
\newcommand{\ycite}[1]{\citeyear{#1}}

\begin{opening}
\title{Search for double degenerate pro\-gen\-i\-tors of 
supernovae type Ia
with SPY}

\author{R.~Napiwotzki}
\author{H.~Drechsel}
\author{U.~Heber}
\author{C.~Karl}
\author{E.-M.~Pauli}
\institute{Dr.~Remeis-Sternwarte, Astronom.\ Institut, Universit\"at 
        Erlangen-N\"urnberg, Sternwartstr.~7, 96049 Bamberg, Germany}
\author{N.~Christlieb}
\author{H.-J.~Hagen}
\author{D.~Reimers}
\institute{Hamburger Sternwarte, Universit\"at Hamburg, Gojenbergsweg 112, 
  21029 Hamburg, Germany}
\author{D.~Koester}
\author{S.~Moehler}
\institute{Institut f\"ur Theoretische Physik und Astrophysik, 
  Universit\"at Kiel, 24098 Kiel, Germany}
\author{D.~Homeier}
\institute{Department of Physics \& Astronomy,
        University of Georgia, Athens, GA\,30602-2451, USA}
\author{B.~Leibundgut}
\author{A.~Renzini}
\institute{European Southern Observatory, Karl-Schwarzschild-Str.~2, 
  85748 Garching, Germany}
\author{T.R.~Marsh}
\institute{University of Southampton, Department of Physics \& Astronomy, 
  Highfield, Southampton S017 1BJ, UK}
\author{G.~Nelemans}
\institute{Institute of Astronomy, Madingley Road, Cambridge CB3~0HA, UK}
\author{L.~Yungelson}
\institute{Institute of Astronomy of the Russian Academy of Sciences, 
  48 Pyatnitskaya Str., 109017 Moscow, Russia}
\end{opening}

\begin{document}


\begin{abstract}
We report on a large survey for double degenerate (DD)
binaries as potential progenitors of type Ia supernovae with the UVES
spectrograph at the ESO VLT (ESO {\bf S}N\,Ia {\bf P}rogenitor
surve{\bf Y} -- SPY).
\end{abstract}


Supernovae of type Ia (SN\,Ia) play an outstanding role for our understanding
of galactic evolution and the determination of the extragalactic distance
scale.  However, the nature of their progenitors is still unknown (e.g.\ Livio
\ycite{Liv00}).  In the so-called double degenerate (DD) scenario (Iben \&
Tutukov \ycite{IT84}) two white dwarfs with a mass exceeding the Chandrasekhar
limit merge.
Several systematic radial velocity ({\bf RV}) searches for DDs have been
undertaken starting in the mid 1980's 
checking a total of $\approx 200$ white dwarfs RV for
variations (cf.\ Marsh \ycite{Mar00} and references
therein),
but have failed to reveal any massive, short-period DD progenitor 
of SN\,Ia. 
However, this is not unexpected, as theoretical simulations suggest that
 only a few percent of all DDs are potential SN\,Ia progenitors
(Iben, Tutukov \& Yungelson \ycite{ITY97}; Nelemans et al.\ \ycite{NYP01}). 

In order to perform a definitive test of the DD scenario we have
embarked on a large spectroscopic survey  of 1500 white dwarfs 
(ESO {\bf S}N \,Ia
{\bf P}rogenitor surve{\bf Y} -- SPY). 
SPY will overcome the main limitation of all efforts so far to detect
DDs that are plausible SN~Ia precursors: the samples of surveyed
objects were too small.  
Spectra were taken with the high-resolution 
UV-Visual Echelle Spectrograph (UVES) of
the UT2 telescope (Kueyen) of the ESO VLT in service mode. 
Our instrument setup 
provides nearly complete spectral coverage from 3200\,\AA\ to 
6650\,\AA\
with a resolution $R=18500$ (0.36\,\AA\ at H$\alpha$). 
Due to the nature of
the project, two spectra at different, ``random'' epochs separated 
by at least one day are observed.

ESO provides a data reduction pipeline for UVES, which
 formed the basis for our first selection
of DD candidates. A careful re-reduction of the spectra is in progress.
Differing from previous surveys we use a correlation procedure to 
determine RV shifts of the observed spectra 
(cf.\ Napiwotzki et al.\ \ycite{NCD01}).
We routinely measure RVs with an accuracy of $\approx 2$\,\kms\
or better, therefore running only a very small risk of missing a merger
precursor, which have orbital velocities of 150\,\kms\ or higher.

\paragraph{Results.}
We have analyzed spectra of 558 white dwarfs and pre-white dwarfs
taken during the first two years
of the SPY project and detected 90 new DDs, 13 are double-lined systems (only 6
were known before). 
SPY is the first RV survey which performs a systematic
investigation of both classes of white dwarfs: DAs {\it and} non-DAs. 
Our observations have already
increased the DD sample by a factor of five.  After completion, a final sample
of $\approx$200 DDs is expected.

\begin{table}
\caption{Fraction of RV variable stars in the current SPY sample for 
different spectral classes. WD+dM denotes systems for which a
previously unknown cool companion is evident from the red spectra
(not included in the DA/non-DA entries).}
\label{t:DDs}

\begin{tabular}{l|r|r|r}
Spectral type   &total  &RV variable    &detection rate\\ \hline
All DDs    &558    &90     &16\%\\
non-DA (DB,DO,DZ)       &72     &10      &14\%    \\
WD+dM   &30     &14      &47\%   \\
\end{tabular}
\end{table}

Follow-up observations of this sample are mandatory to 
determine periods and white dwarf parameters
and find potential SN\,Ia progenitors among the candidates. 
Good statistics of a large DD sample will also set stringent
constraints on the evolution of close binaries, which will
dramatically improve our understanding of this phase of stellar
evolution.  
Starting in 2001 follow-up observations have been carried out with VLT 
and NTT of ESO as 
well as with the 3.5\,m telescope of the Calar Alto observatory/Spain
and the INT (Napiwotzki et al.\ \ycite{NCD01}, \ycite{NEH01}, \ycite{NKN02};
Karl et al., these proceedings). 
Our sample includes many short period binaries, several with masses closer to
the Chandrasekhar limit than any system known before.  
During our follow-up observations we have detected a very promising
potential SN\,Ia precursor candidate. However, some additional spectra
are necessary to verify our RV curve solution. Results will be
reported elsewhere.

\begin{figure}
\epsfxsize=10cm
\epsfbox[-30 40 740 550]{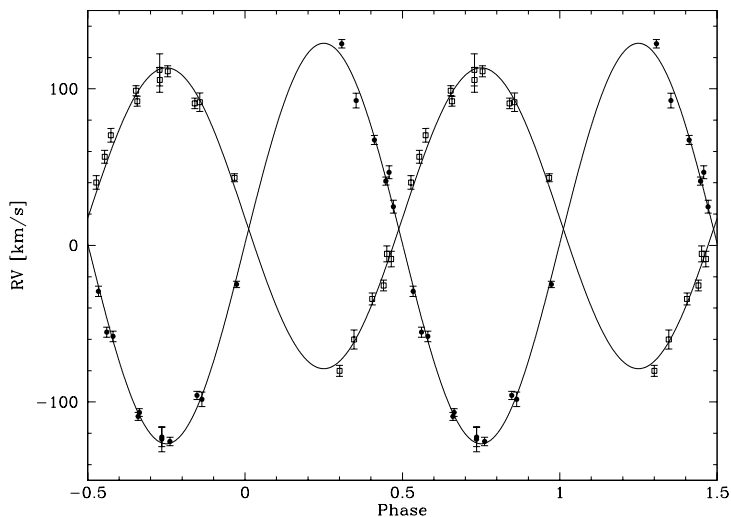}
\caption{Measured RVs as a function of orbital phase and
fitted sine curves for HE\,1414-0848. Filled circles/open rectangles 
 indicate the less/more massive components A and B. 
Note the difference of the
``systemic velocities'' $\gamma_0$ between both components caused by
gravitational redshift.}
\label{f:HE1414rv}
\end{figure}

Exemplary for other double-lined systems we discuss here the 
DA+DA system HE\,1414$-$0848.
The orbital period of
$P = 12^{\mathrm{h}} 25^{\mathrm{m}} 44^{\mathrm{s}}$ and
semi-amplitudes of 127\,km\,s$^{-1}$ and 96\,km\,s$^{-1}$ are derived
for the individual components (Napiwotzki et al.\ \ycite{NKN02}).  
RV curves for both
components are displayed
in Fig.~\ref{f:HE1414rv}. 
The ratio of velocity amplitudes is directly related to the
mass ratio of both components. Additional information comes from the 
mass dependent gravitational redshift. 
The difference in gravitational redshift corresponds to the apparent
difference of ``systemic velocities'' of both components, as derived from the
RV curves (Fig.~1).  Only one set of individual white dwarf masses fulfills
the constraints given by both the amplitude ratio and redshift difference (for
a given mass-radius relation).  We estimate the masses of the individual
components with this method to be $0.55M_\odot$ and $0.71M_\odot$ for A and B,
respectively.
This translates into $\log g = 7.92$ and 8.16, respectively.
 
Another estimate of the white dwarf parameters is available from a model
atmosphere analysis of the combined spectrum. We have developed a new tool
({\sc FITSB2}), which performs a spectral analysis of both components of a
double-lined system. The fit is performed on all available spectra, covering
different orbital phases simultaneously.  We fitted temperatures and gravities
of both components of HE\,1414$-$0848 (the mass ratio fixed at the accurate
value derived from the RV curve).  The results are $T_{\mathrm{eff}}$/$\log g$
= 8380\,K/7.83 and 10900\,K/8.14 for A and B, which are in good agreement with
the $\log g$ values predicted from the analysis of the RV curve.  The total
mass of the HE\,1414$-$0848 system is $1.26M_\odot$, only 10\% below the
Chandrasekhar limit.  The system will merge due to loss of angular momentum
via gravitational wave radiation after two Hubble times.

\paragraph{Spin-off results.} SPY produces an immense, unique
sample of very high resolution white dwarf spectra. It will allow us for
the first time to tackle many longstanding questions on a firm
statistical basis.  Among those are the mass distribution of white dwarfs
(Koester et al.\ \ycite{KNC01}), 
kinematical properties of the white dwarf population (Pauli, these
proceedings), surface compositions,
luminosity function, rotational velocities, and detection of weak
magnetic fields. 
A more detailed description of ongoing spin-off activity
is given in Napiwotzki et al.\ \scite{NCD01}. 
Members of the community interested in spin-off opportunities are invited
to participate in the exploitation of the SPY sample.


\end{document}